\documentclass[prc,twocolumn,showpacs]{revtex4}
\usepackage{graphicx}
\usepackage{float}
\def\rbot{r_{\scriptscriptstyle{\bot}}}
\newcommand{\D}{\displaystyle}

\begin{document}
\title {Reflection asymmetric relativistic mean field approach and its application to the octupole deformed nucleus $^{226}$Ra}
\author{
L. S. Geng$^{1,2}$,  J. Meng$^{1,2,3}$, and H. Toki$^{4}$}
\affiliation{
$^1$School of Physics, Peking University, Beijing 100871\\
$^2$Institute of Theoretical Physics, Chinese Academy of
Sciences, Beijing 100080\\
$^3$Center of Theoretical Nuclear Physics, National Laboratory of
       Heavy Ion Accelerator, Lanzhou 730000\\
$^4$Research Center for Nuclear Physics(RCNP), Osaka
University, Ibaraki 567-0047, Japan}
\date{\today}

\begin{abstract}
A Reflection ASymmetric Relativistic Mean Field (RAS-RMF) approach
is developed by expanding the equations of motion for both the
nucleons and the mesons on the eigenfunctions of the two-center
harmonic-oscillator potential. The efficiency and reliability of the
RAS-RMF approach are demonstrated in its application to the
well-known octupole deformed nucleus $^{226}$Ra and the available
data, including the binding energy and the deformation parameters,
are well reproduced.
\end{abstract}

\pacs{21.60.-n; 21.60.Jz; 21.10.Ft; 21.10.Gv}

\maketitle

The relativistic mean field (RMF) model is one of the most
successful microscopic models in nuclear physics. It incorporates
from the beginning very important relativistic effects, such as the
existence of two types of potentials (Lorentz scalar and
four-vector) and the resulting strong spin-orbit
interaction~\cite{Ring96,Meng05}. During the past two decades, it
has received wide attention due to its successes in describing many
nuclear phenomena for stable nuclei, exotic nuclei, as well as
supernovae and neutron stars \cite{Ring96,Meng05}. Recently, a
systematic theoretical study of over 7000 nuclei has been performed
in the RMF model and overall good agreements with existing data were
obtained \cite{Geng05PTP}. To solve the RMF equations for finite
systems, in particular deformed systems, the basis expansion method
is most widely employed \cite{Ring97,Hirata96,Meng05Nucl-th}, though
other alternatives exist \cite{Rutz95,Meng98npa,Poschl98,Zhou03prc}.
The basis expansion method is very efficient, and, more importantly,
very intuitive.  Due to the limitation of the conventional
harmonic-oscillator potential (spherical, axial, or triaxial),
reflection symmetry is always assumed in these methods
\cite{Ring97,Hirata96}.

On the other hand, as early as in the 1950s, it was realized that
some nuclei may have a shape asymmetric under reflection, such as a
pear shape \cite{Lee57}, with the observation of low-lying $1^-$
states in the actinide nuclei~\cite{Asaro53,Stephens54,Stephens55}.
Microscopically, this is attributed to the coupling between
single-particle states which differ by $\Delta\ell=3$ and $\Delta
j=3$. These states lie close to each other and to the Fermi surface
at proton and neutron numbers near 34, 56, 88 and neutron number
near 134, where octupole correlations are expected to be the
strongest \cite{Butler96}. Octupole deformation increases the
nuclear binding energy by only a few MeV for most cases, but it is
essential to explain a lot of experimental observations
\cite{Butler96}, such as the appearance of parity doublets
\cite{Chasman80}. It can also lower the second fission-barrier of
heavy and superheavy nuclei and explain the asymmetric mass
distribution observed in the fission of $^{240}$Pu \cite{Rutz95}.

In the present work, we develop a Reflection ASymmetric Relativistic
Mean Field (RAS-RMF) approach so that nuclei having
reflection-asymmetric shapes can be studied (We nevertheless assume
axial symmetry). To achieve this and to follow the general principle
of the basis expansion technique~\cite{Vautherin73,Ring97}, we
employ the eigenfunctions of the two-center harmonic-oscillator
(TCHO) potential, which was traditionally used in the two-center
shell model (TCSM) by the Frankfurt group~\cite{Holzer69,Maruhn72}.
The TCHO basis has been widely used in studies of fission, fusion,
heavy-ion emission, and various cluster phenomena
studies~\cite{Greiner94}. Its applications in microscopic mean-field
models, however, are very rare except a few in non-relativistic
Hartree-Fock models in the 1970s~\cite{Passler73,Dreizler74}.

\begin{table*}[htpb]
\vspace{0.2cm}
\setlength{\tabcolsep}{0.6 em} \caption{Convergences of the
RAS-RMF calculations versus the number of shells $N(=N_F=N_B)$ for the ground-state properties of $^{16}$O
and $^{208}$Pb, in comparison with those of the axial deformed RMF
calculations~\cite{Ring97}.}
\begin{center}
\begin{tabular}{c|c|c|c|c|c|c|c|c|c|c|c|c}
\hline\hline
\multicolumn{13}{c}{$^{16}$O}\\
\hline
 &\multicolumn{6}{c|}{Binding energy per nucleon (MeV)}&\multicolumn{6}{c}{Charge radius   (fm)}\\
 \hline
$N$ &10&12&14&16&18&20&10&12&14&16&18&20\\
 \hline
Axial deformed RMF&8.052&8.050&8.051&8.051&8.051&8.051&2.724&2.725&2.727&2.727&2.728&2.728\\
RAS-RMF &8.052&8.050&8.051&8.051&8.051&8.051&2.724&2.725&2.727&2.727&2.728&2.728\\
\hline\hline
\multicolumn{13}{c}{$^{208}$Pb}\\
\hline
 &\multicolumn{6}{c|}{Binding energy per nucleon (MeV)}&\multicolumn{6}{c}{Charge radius (fm)}\\
 \hline
$N$ &10&12&14&16&18&20&10&12&14&16&18&20\\
 \hline
Axial deformed RMF&7.837&7.896&7.896&7.888&7.887&7.886&5.515&5.517&5.509&5.512&5.516&5.515\\
RAS-RMF &7.852&7.891&7.891&7.886&7.886&7.884&5.523&5.510&5.510&5.516&5.517&5.516\\
 \hline\hline
\end{tabular}
\end{center}
\end{table*}
\begin{table*}[htpb]
\setlength{\tabcolsep}{0.6 em} \caption{ Ground-state properties of
$^{226}$Ra, including  the binding energy $E_B$, the quadrupole,
octupole, and hexadecapole deformation parameters for neutron,
proton(charge), and matter density distributions ($\beta_{\ell n}$,
$\beta_{\ell p(c)}$ and $\beta_\ell$ with $\ell=2$, 3, and 4)
obtained from the RAS-RMF and axial deformed RMF
 calculations, in comparison with the data. }
\begin{center}
\begin{tabular}{c|ccc|ccc|ccc|c}
\hline\hline
 &$\beta_{2n}$&$\beta_{2p(c)}$&$\beta_2$&$\beta_{3n}$&$\beta_{3p(c)}$&$\beta_3$
 &$\beta_{4n}$&$\beta_{4p(c)}$&$\beta_4$& $E_B$ [MeV]\\
 \hline
 Axial deformed RMF & 0.19 & 0.18 & 0.19 & -- & -- & -- & 0.14 & 0.12 & 0.13 & 1729.1\\
 RAS-RMF  & 0.21 & 0.20 & 0.21 & $-0.16$ & $-0.15$ & $-0.16$ & 0.17 & 0.15 & 0.16 & 1731.8\\
  Exp.&     & 0.20 &      &      & -0.13 &      &     &       &      &  1731.6 \\
 \hline\hline
\end{tabular}
\end{center}
\end{table*}
The RMF model used here is the standard one and its detailed
reviews can be found in Refs.~\cite{Ring96,Meng05}; therefore, only a
brief introduction is provided in the following. The adopted
Lagrangian density is
\begin{eqnarray}
\mathcal{L} &=& \bar \psi (i\gamma^\mu\partial_\mu -M) \psi\nonumber \\
&+&\,\frac{1}{2}\partial_\mu\sigma\partial^\mu\sigma -\frac{\D 1}{\D
2}m_{\sigma}^{2} \sigma^{2}- \frac{\D 1}{\D 3} g_{2} \sigma^{3} -
\frac{\D 1}{\D 4}g_{3}\sigma^{4} - g_{\sigma}\bar\psi
\sigma \psi \nonumber\\
&-&\frac{\D 1}{\D 4}\Omega_{\mu\nu}\Omega^{\mu\nu}+\frac{\D 1}{\D
2}m_\omega^2\omega_\mu\omega^\mu +\frac{\D 1}{\D
4}c_4(\omega_\mu\omega^\mu)^2-g_{\omega}\bar\psi
\gamma^\mu \psi\omega_\mu\nonumber\\
 &-& \frac{\D 1}{\D 4}{R^a}_{\mu\nu}{R^a}^{\mu\nu} +
 \frac{\D 1}{\D 2}m_{\rho}^{2}
 \rho^a_{\mu}\rho^{a\mu}
     -g_{\rho}\bar\psi\gamma_\mu\tau^a \psi\rho^{\mu a} \nonumber\\
      &-& \frac{\D 1}{\D 4}F_{\mu\nu}F^{\mu\nu} -e \bar\psi
      \gamma_\mu\frac{\D 1-\tau_3}{\D 2}A^\mu
      \psi,
\end{eqnarray}
where all symbols have their usual meanings. Using the classical
variational principle, one can obtain the Dirac equation for the
nucleons and Klein-Gordon equations for the mesons
\cite{Ring96,Meng05}. To solve these equations, we employ the basis
expansion method, which has been widely used in both
non-relativistic and relativistic mean-field models~
\cite{Vautherin73,Ring97}. For axial-symmetric reflection-asymmetric
systems, the spinors $f_i^\pm$ and $g_i^\pm$ can be expanded in
terms of the eigenfunctions of the TCHO potential
\begin{equation}\label{potential}
V(\rbot,z)=\frac{1}{2}M\omega^2_\bot\rbot^2+\left\{\begin{array}{l}
\frac{1}{2}M\omega_1^2(z+z_1)^2,\quad z<0\\
\frac{1}{2}M\omega^2_2(z-z_2)^2,\quad z\ge0\end{array}\right.,
\end{equation}
where $M$ is the nucleon mass,  $z_1$ and $z_2$ (real, positive)
represent the distances between the centers of the spheroids and
their intersection plane, and $\omega_1$($\omega_2$) are the
corresponding oscillator frequencies for $z<0$ ($z\ge0$)~
\cite{Holzer69,Maruhn72,Mirea96}.

Imposing the assumption of volume-surface
conservation~\cite{Holzer69,Maruhn72,Mirea96}, the TCHO basis can be
completely specified by three parameters: $\delta_2$, the quadrupole
deformation parameter of the $z>0$ half spheroid which defines the
ratio of $\omega_2$ and $\omega_\bot$ through
$\frac{\omega_2}{\omega_\bot}=
\exp(-\frac{3}{2}\sqrt{\frac{5}{4\pi}}\delta_2)$~\cite{Ring97};
$\delta_3$, which defines the ratio of $\omega_1$ and $\omega_2$
through $\delta_3=\frac{\omega_1}{\omega_2}$; and $\Delta
z=z_1+z_2$, the distance between the two centers. Here, the
spherical harmonic-oscillator parameters are chosen as
$\hbar\omega_0=41A^{-1/3}$ and $R_0=1.2A^{1/3}$ as in
Refs.~\cite{Holzer69,Maruhn72,Vautherin73,Ring97}. Since the primary
purpose of the present work is to extend the RMF model
 to reflection-asymmetric systems, only the $\delta_3$
degree of freedom of the TCHO basis is explored while the $\Delta
z$ degree of freedom will be left for future studies, i.e. we set $\Delta
z\approx0$.

For $\delta_3=1.0$ and $\Delta z\approx0$, the TCHO potential
reduces to the reflection-symmetric harmonic-oscillator potential
and the RAS-RMF method becomes equivalent to the axial deformed RMF
approach developed in Ref.~\cite{Ring97}. In Table I, the results
obtained from the RAS-RMF calculations are compared with those
obtained from the axial deformed RMF calculations for $^{16}$O and
$^{208}$Pb. The mean-field force employed is NL3~\cite{Lala97}.
Using other forces, such as NL1 \cite{Reinhard86}, does not
influence our conclusion here. As expected, the agreement between
these two calculations is remarkable. This demonstrates the
equivalence of these two codes in the special case. Table I also
shows that the results already converge at $N(=N_F=N_B)=16$ even for
$^{208}$Pb. In the following, unless indicated otherwise, numerical
calculations will be performed by using $N=16$. For
$\delta_3\ne1.0$, one should note that the eigenvalues of
$\phi_\nu(z)$, $\nu_1$ and $\nu_2$, are not integers anymore.
Therefore, in this case, the dimension of the basis (denoted by
$\tilde{N}$) is chosen to be the same as that at $\delta_3=1.0$
(denoted by $N$).

In the following, we apply the RAS-RMF approach to the nucleus
$^{226}$Ra, which has been predicted to have a reflection-asymmetric
ground state by various theoretical models (see Ref.~\cite{Rutz95}
and references therein). The purpose is to demonstrate the validity
of the RAS-RMF approach. For the mean-field channel, two effective
forces have been used, i.e NL3 \cite{Lala97} and NL1
\cite{Reinhard86}.  The numerical results obtained from these two
forces turn out to be quite similar for $^{226}$Ra, in particular
for the deformation parameters. Therefore, only the numerical
results obtained with NL1 are presented.  For the pairing channel,
the constant gap approach with $ \Delta=11.2\,\mbox{MeV}/\sqrt{A}$
is used.

\begin{figure}[htpb]
 \centering
\includegraphics[scale=0.36]{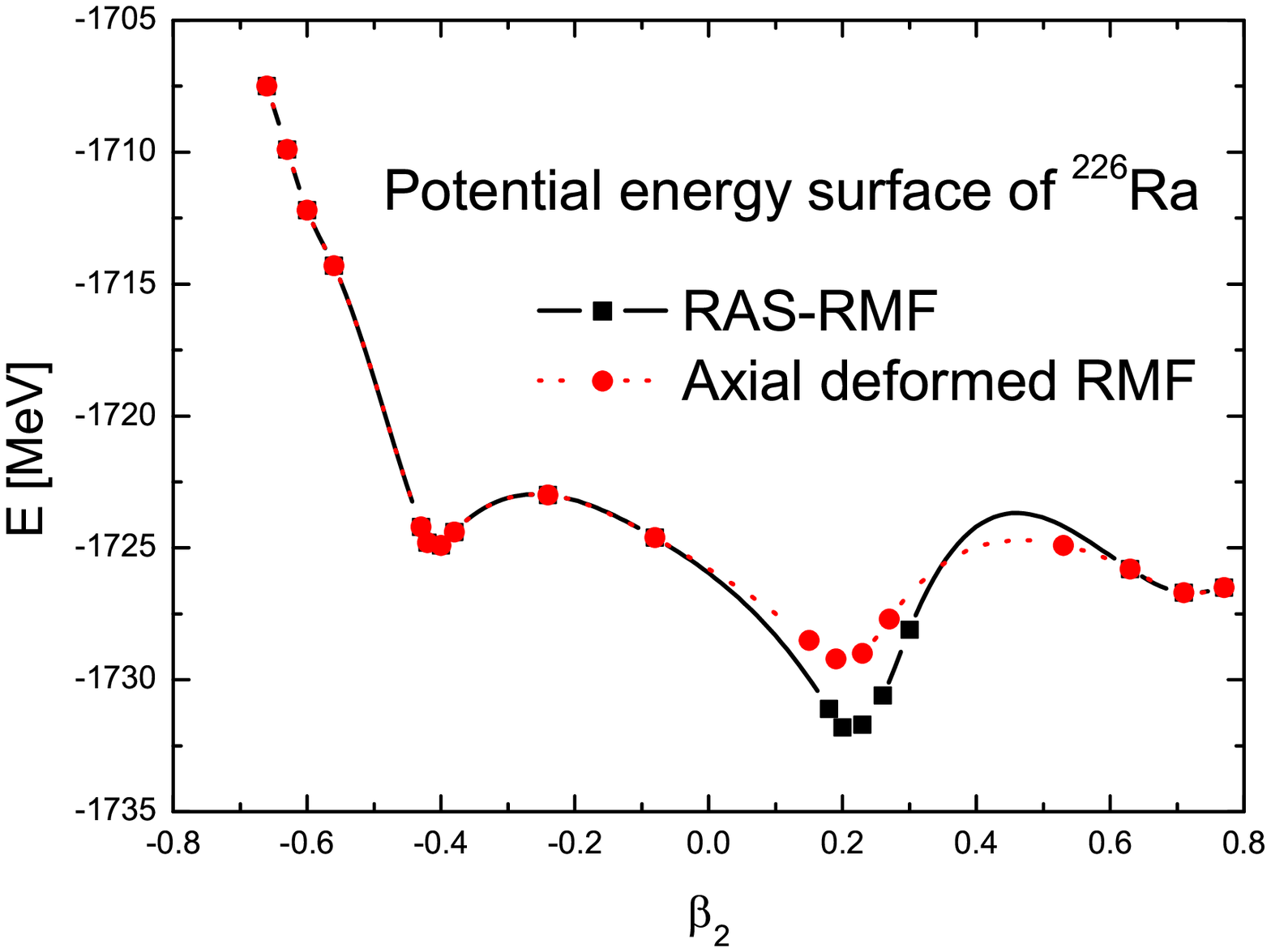}
 \includegraphics[scale=0.36]{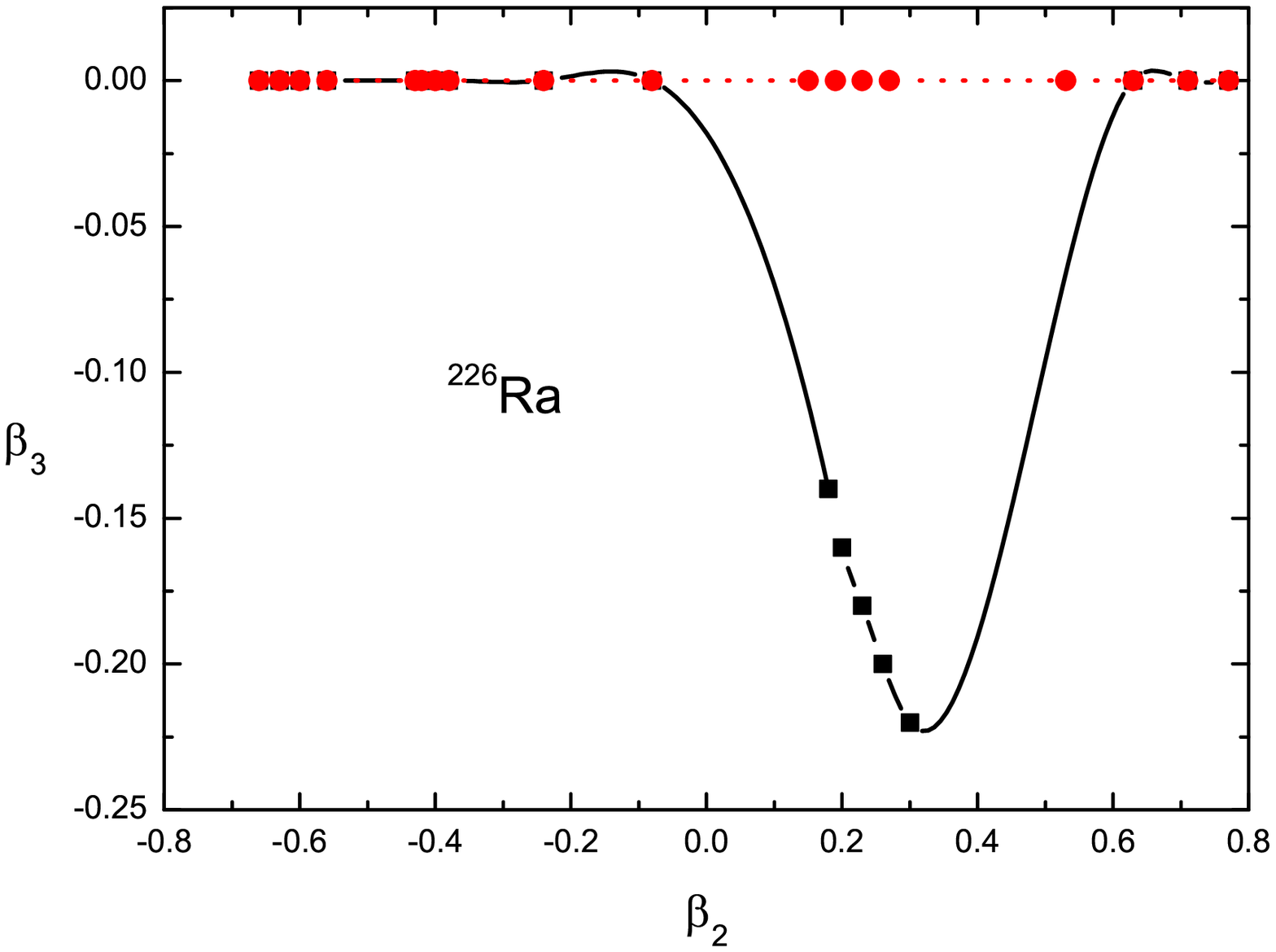}
\caption{Upper panel: the potential energy surfaces of $^{226}$Ra as
versus the quadrupole deformation parameter $\beta_2$; lower panel:
the octupole deformation parameter $\beta_3$ of $^{226}$Ra as a
function of $\beta_2$ obtained from the RAS-RMF calculations.}
\end{figure}

\begin{figure*}[htpb]
 \centering
\includegraphics[scale=0.38]{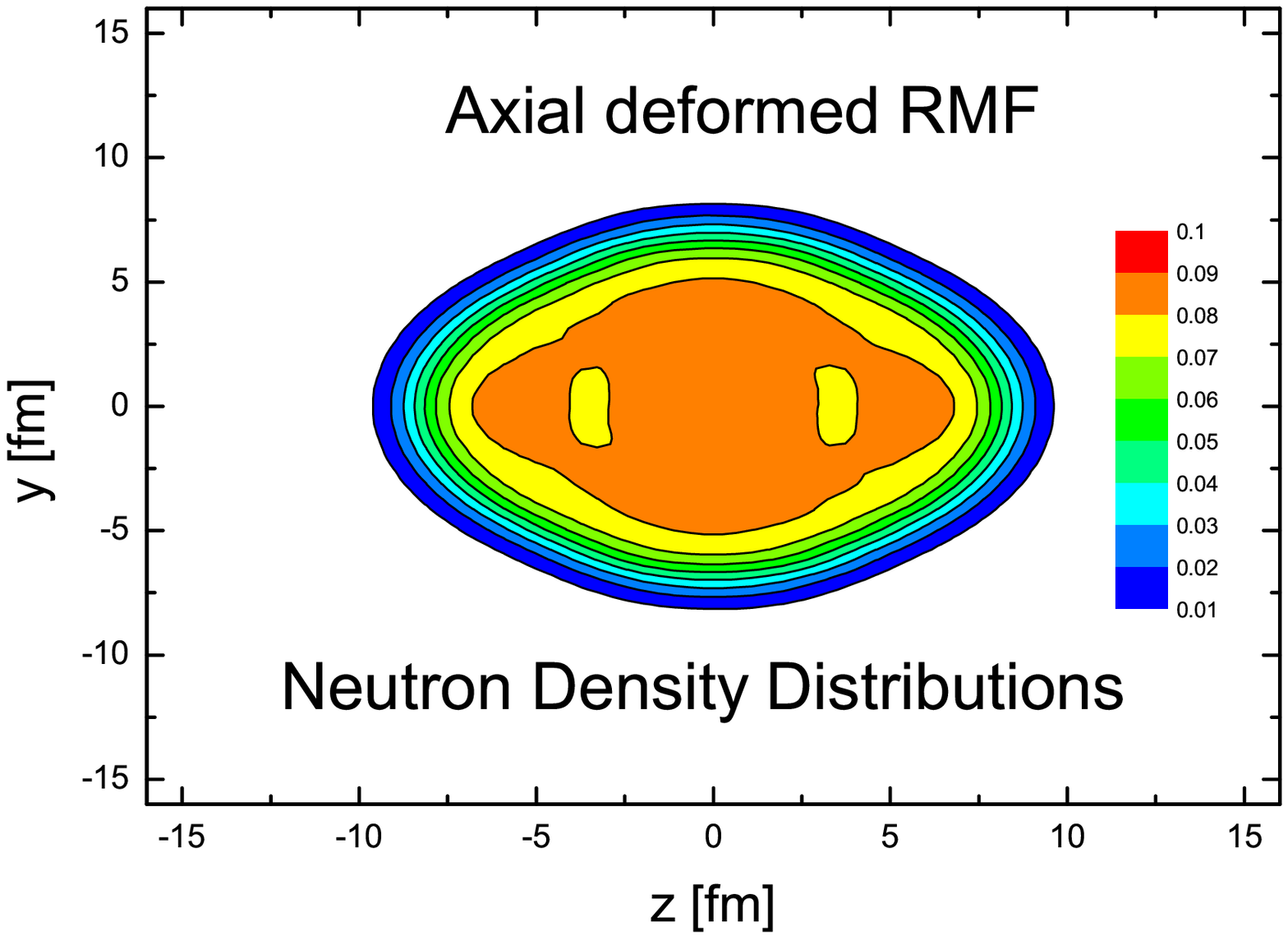}%
\includegraphics[scale=0.38]{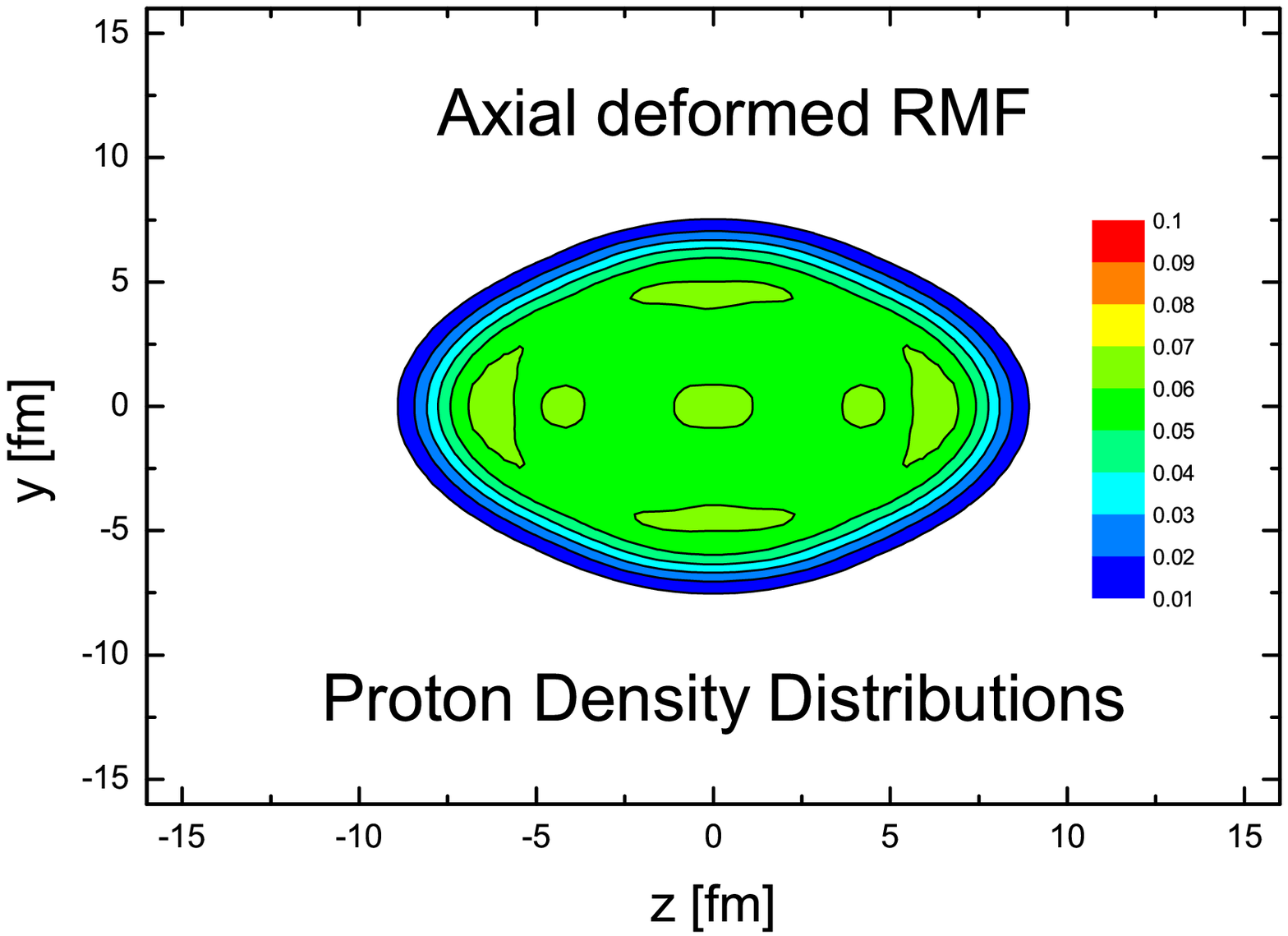}
\includegraphics[scale=0.38]{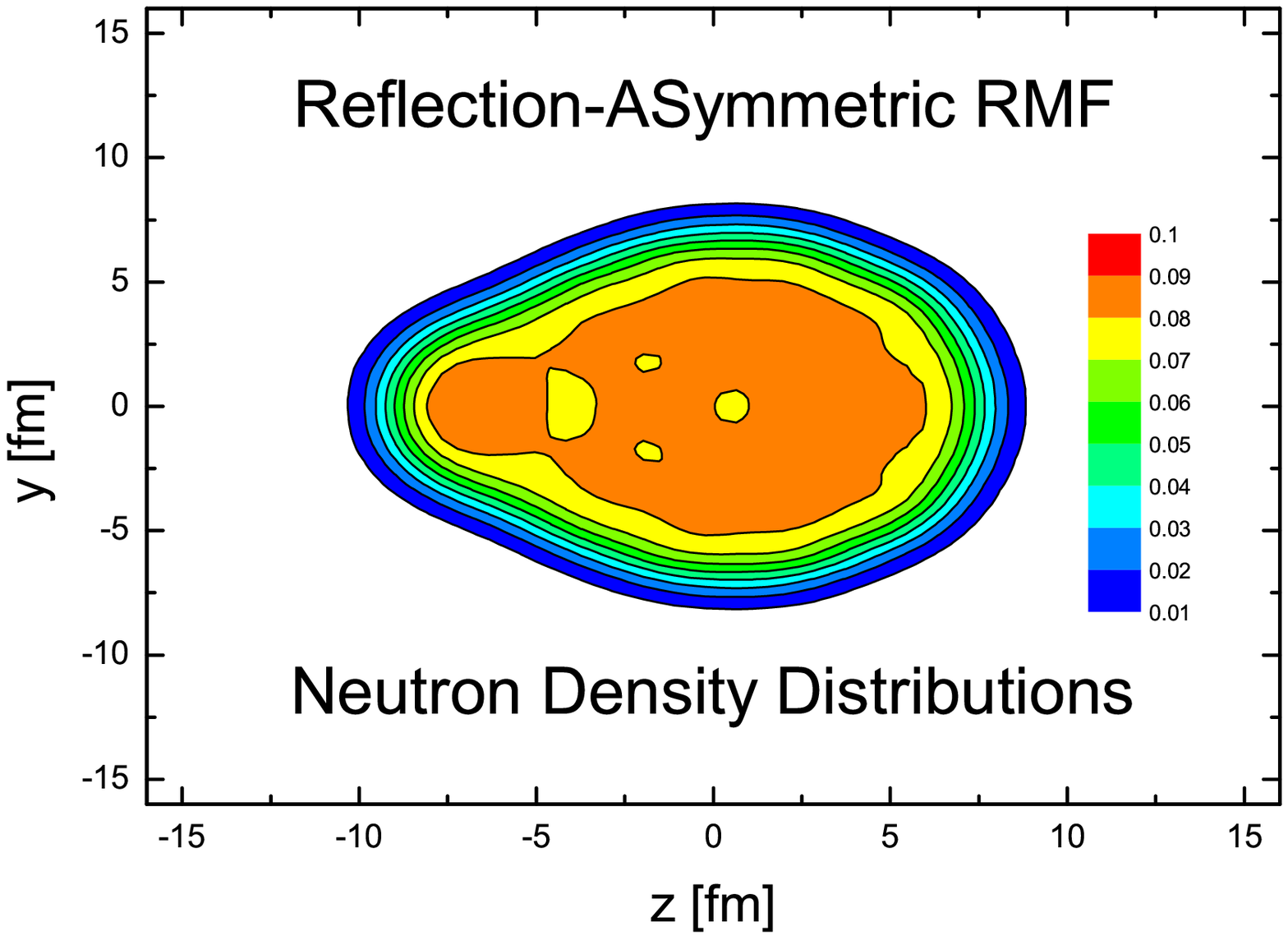}%
\includegraphics[scale=0.38]{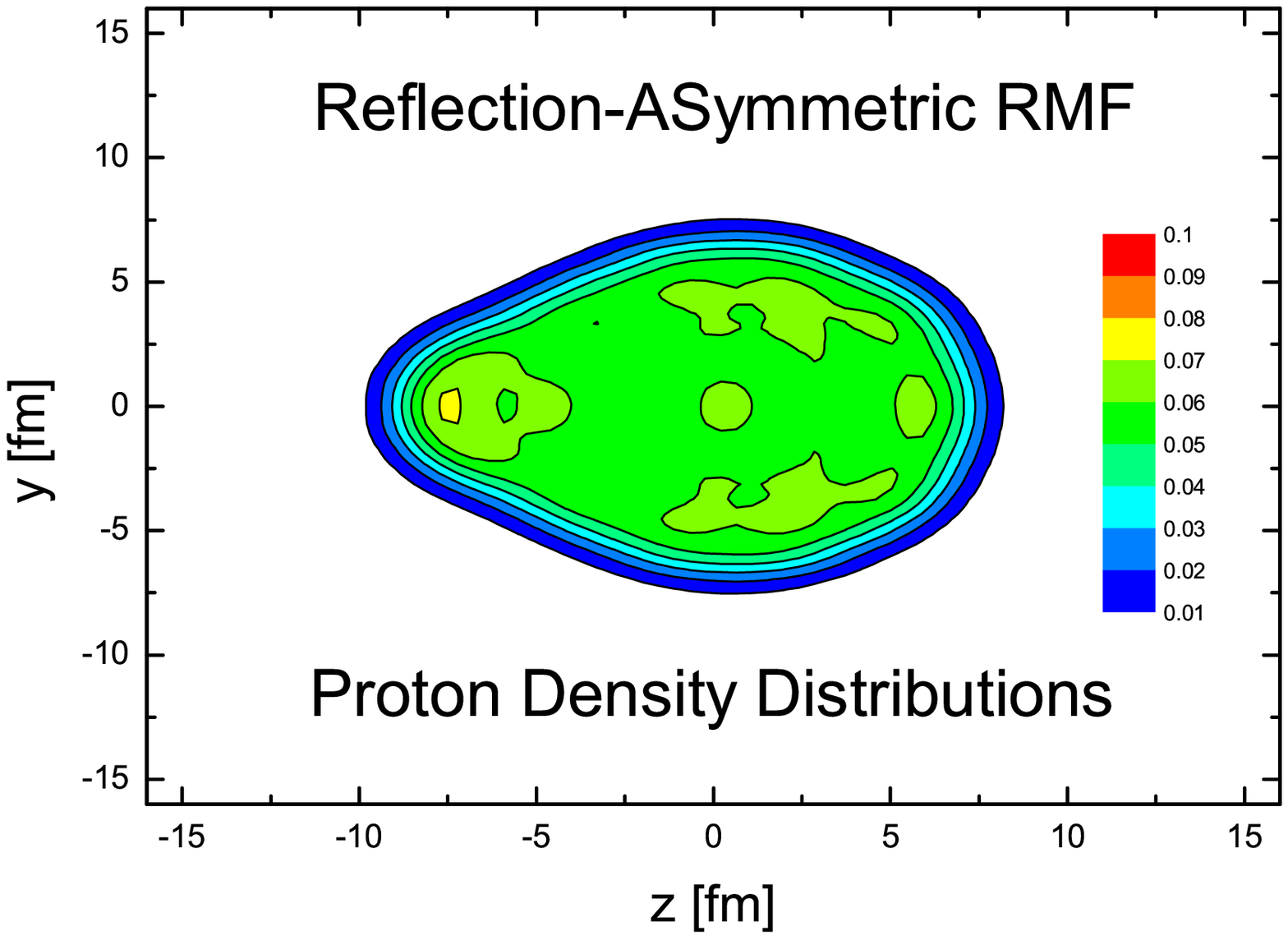}
\caption{Neutron and proton density distributions of the ground
state of $^{226}$Ra versus $z$ and $y$ on the $x=0$ plane. Upper
panel: results obtained from the axial deformed RMF calculations;
lower panel: results obtained from the RAS-RMF calculations.}
\end{figure*}
On the upper panel of Fig.~1, the potential energy surfaces of
$^{226}$Ra obtained from RAS-RMF and axial deformed RMF calculations
are plotted as functions of the quadrupole deformation parameter
$\beta_2$. These are obtained by a quadratic constraint on the
quadrupole moment $\langle Q_2\rangle=\langle
r^2Y_{20}\rangle$~\cite{Flocard73} while the octupole deformations
have been left free to adjust themselves to the minimum
configuration. It is easily seen that the asymmetric  ground state
is lower than the symmetric ground state by roughly 3\,MeV while the
asymmetric barrier at $\beta_2\approx0.5$ is slightly higher than
the corresponding symmetric barrier. For the other parts of the
potential energy surfaces, these two calculations agree with each
other. It should be mentioned that all these are in coincidence with
the results of Ref.~\cite{Rutz95}. Naturally, one would think that
the above differences originate from the octupole degree of freedom,
which is in fact the case as can be seen from the lower panel of
Fig.~1.

In Table II, the ground-state properties of $^{226}$Ra obtained from
RAS-RMF and axial deformed RMF calculations are tabulated. It is
clearly seen that (i) the asymmetric ground state is more bound by
roughly 2.7\,MeV than the symmetric ground state, (ii) the
quadrupole and hexadecapole deformations obtained from the RAS-RMF
calculations are slightly larger than those from the axial deformed
RMF calculations due to the emergence of octupole deformation, and
(iii) the slight difference between the neutron octupole deformation
and the proton octupole deformation would generate a non-vanishing
static electric dipole moment \cite{Rutz95}. The experimental
binding energy of $^{226}$Ra is 1731.6\,MeV \cite{Audi03}, which
agrees well with our RAS-RMF prediction, 1731.8\,MeV (see Table II).
Experimentally, the charge quadrupole and octupole deformation
parameters can be  extracted from the experimental $B(E2)\uparrow$
\cite{Raman01} and $B(E3)\uparrow$ \cite{Butler96} values  by
\begin{equation}
\beta_{2c}=\frac{4\pi}{3ZR_0^2}\left[\frac{B(E2)\uparrow}{e^2}\right]^{1/2},
\,
\beta_{3c}=\frac{4\pi}{3ZR_0^3}\left[\frac{B(E3)\uparrow}{e^2}\right]^{1/2}.
\end{equation}
From  $B(E2)\uparrow$ of  5.1514 $b^2 e^2$ \cite{Raman01} and
$B(E3)\uparrow$ of 1.1011 $b^3e^2$ \cite{Wo93} for $^{226}$Ra, one
can deduce the corresponding $\beta_2$ and $\beta_3$ to be 0.20 and
0.13, respectively, which agree well with our predictions as what
can be seen from Table II. It should be noted that our calculations
are symmetric for $\beta_3>0$ and $\beta_3<0$. Therefore, in Table
II, the experimental $\beta_3$ is chosen to have a negative sign.

The nuclear shape can be visualized from its density distributions.
In Fig.~2, the neutron and proton density distributions of the
ground state of $^{226}$Ra obtained from the axial deformed RMF
calculations and those obtained form our RAS-RMF calculations are
plotted as functions of $z$ and $y$ on the $x=0$ plane. It is easily
seen that the ground state in the axial deformed RMF calculations
corresponds to a nucleus of a spheroid shape while the ground state
in our RAS-RMF calculations corresponds to a nucleus of a pear
shape. It is noticed that the neutron density at the center of the
nucleus is larger than the corresponding proton density due to the
larger number of neutrons as compared to that of protons.

In summary, we have solved the RMF equations for axial-symmetric
reflection-asymmetric systems, i.e. we have developed the RAS-RMF
approach. The numerical details of this approach are presented. It
is then applied to investigate the ground-state properties of
$^{226}$Ra and good agreements with existing data are obtained. The
RAS-RMF approach, therefore, provides another microscopic framework
to study nuclei with intrinsic reflection-asymmetric shapes and the
rich nuclear phenomena related \cite{Butler96}. A nonzero atomic
electric-dipole moment (EDM) would signify time-reversal violation
from outside the Standard Model; while EDMs are enhanced in atoms
that have octupole deformed nuclei, such as
$^{225}$Ra~\cite{Dobaczewski05}. This can be studied in the future
within the RAS-RMF framework. This approach with the $\Delta z$
degree of freedom may also be useful in studies of largely deformed
systems, for example, those during fission or fusion.

We thank E. G. Zhao, S.-G. Zhou, and H. F. L\"{u} for stimulating
discussions. This work is supported in part by the National Natural
Science Foundation of China under Grant Nos. 10435010, 10221003, and
the Doctoral Program Foundation from the Ministry of Education in
China.

\end{document}